# High-resolution three-dimensional imaging of topological textures in single-diamond networks

D. Karpov[1,§], K. Djeghdi[2], M. Holler[1], S. Narjes Abdollahi[2,†], K. Godlewska[2], C. Donnelly[3], T. Yuasa[4,¶], H. Sai[4,‡], U. B. Wiesner[4,5], B. D. Wilts[2,6], U. Steiner[2], M. Musya[7], S. Fukami[7,8,9,10,11], H. Ohno[7,8,9,10], I. Gunkel[2], A. Diaz[1] and J. Llandro[7,8,] *

[1] Paul Scherrer Institute, Forschungsstrasse 111, 5232 Villigen, Switzerland.
[2] Adolphe Merkle Institute, Chemin des Verdiers 4, 1700 Fribourg, Switzerland.
[3] Max Planck Institute for Chemical Physics of Solids, Nöthnitzer Str. 40, 01187 Dresden, Germany.
[4] Department of Materials Science and Engineering, Cornell University, Ithaca, NY 14853, USA.
[5] Kavli Institute at Cornell for Nanoscale Science, Cornell University, Ithaca, NY 14853, USA.
[6] Chemistry and Physics of Materials, University of Salzburg, Jakob-Haringer-Straße 2a, 5020 Salzburg, Austria.
[7] Laboratory for Nanoelectronics and Spintronics, Research Institute of Electrical Communication, Tohoku University, 2-1-1 Katahira, Aoba-ku, Sendai 980-8577, Japan.
[8] Center for Science and Innovation in Spintronics, Tohoku University, 2-1-1 Katahira, Aoba-ku, Sendai 980-8577, Japan.
[9] Center for Innovative Integrated Electronic Systems, Tohoku University, 468-1 Aramaki Aza Aoba, Aoba-ku, Sendai 980-0845 Japan.
[10] WPI Advanced Institute for Materials Research, Tohoku University, 2-1-1 Katahira, Aoba-ku, Sendai 980-8577, Japan.
[11] Inamori Research Institute for Science, Kyoto 600-8411, Japan.

§ current address: European Synchrotron Radiation Facility, 71 Av. des Martyrs, 38000 Grenoble, France.
† current address: Department of Chemistry, University of Basel, Mattenstrasse 24a, BPR-1096, 4058 Basel, Switzerland.
¶ current address: Yokkaichi Research Center, JSR Corporation, Yokkaichi, Mie, 510-8552 Japan.
‡ current address: Simpson Querrey Institute for Bionanotechnology, Northwestern University Evanston, IL 60208, USA.
* corresponding author: llandro.justin.b6@tohoku.ac.jp

## Abstract

Highly periodic structures are often said to convey the beauty of nature. However, most material properties are strongly influenced by the defects they contain. On the mesoscopic scale, molecular self-assembly exemplifies this interplay; thermodynamic principles determine short-range order, but long-range order is mainly impeded by the kinetic history of the material and by thermal fluctuations. For the development of self-assembly technologies, it is imperative to characterise and understand the interplay between self-assembled order and defect-induced disorder. Here we used synchrotron-based hard X-ray nanotomography to reveal a pair of extended topological defects within a self-assembled single-diamond network morphology. These defects are morphologically similar to the comet and trefoil patterns of equal and opposite half-integer topological charges observed in liquid crystals and appear to maintain a constant separation across the thickness of the sample, resembling pairs of full vortices in superconductors and other hard condensed matter systems. These results are expected to open new windows to study defect formation in soft condensed matter, particularly in biological systems where most structures are formed by self-assembly.

Introduction

The role of defects depends on their nature. While point defects can alter the local material properties, extended defects involving topology can have far more profound consequences, affecting the global material properties. Topological defects have been found to underpin many known physical phenomena. They can lead to new or enhanced material properties, making understanding their formation important for fundamental physics and next-generation nanotechnology[1]. Research on topological defects spans both soft and hard artificial condensed matter systems. They were first extensively studied in liquid crystals[2] where visible light techniques can be employed. In hard condensed matter topological defects include Dirac chains and emergent magnetic monopoles in spin ices, skyrmions, and vortices and flux tubes in superconductors[3-5]. Recently, topological defects and textures have been found in the actin fibres of single-celled organisms[6], colonies of motile bacteria[7,8], and nacre in bivalves[9] where they play a critical role in development and regeneration.

An outstanding question, therefore, concerns whether mesoscale topological defects can be observed in 'bridge' systems that share biological and condensed matter properties. Soft condensed matter covers synthetic structures, combining macromolecular flexibility and formation kinetics with the periodic ordering of rigid atomic crystals. A convenient pathway to experimentally realising soft condensed matter systems is block copolymer (BCP) self-assembly[10], which offers access to a wide range of morphologies[11] whose properties can be tuned by molar mass and copolymer composition, as well as the processing (annealing) conditions[12]. Among these, 3D ordered continuous networks[13,14] are relatively rare. A particularly well-known example is the gyroid morphology, which has attracted wide interest for the chiral[15,16] and topological[17,18] properties of the single gyroid network of 3-connected nodes formed by triads of nanowire-like struts. The single diamond morphology[13,14] (space group $Fd\bar{3}m$), a tetrahedral network with four struts per node (see the level set model in Figure 1a), has also been predicted to be stable both in linear[19] and branched[20] BCPs. It has attracted enormous interest because of its potential to generate a complete photonic band gap when it features the correct dimensions and refractive indices[20,21]. However, it has proven highly challenging to generate a single diamond network as a well-ordered phase even in additive-laden BCPs[22,23]. We report our observations of the single-diamond morphology templated from a neat linear BCP in a separate publication[24]. In the current work, we observed for the first time mesoscale topological defects in an extended sample of a single-diamond network replicated from a neat BCP (triblock terpolymer) template. Using synchrotron-based hard X-ray nanotomography we imaged a cylindrical sample 8 μm in diameter and 3 μm in height containing a 600 nm thick layer of the single-diamond network. We achieved a 3D spatial resolution of 11.2 nm. This high spatial resolution allowed us to resolve the structure of individual single-diamond unit cells (around 49×75×108 nm$^3$) and analyze long-range order of the network. Long-range order analysis allowed us to identify a pair of topological defects – a "comet"-like and a "trefoil"-like texture – that emerged at the boundaries between diamond domains of different orientations. We confirm the topological nature of the defects through analysis of their winding number and infer their formation mechanism through mapping of distortions in the diamond network. Our analysis suggests that topological defects emerge from the BCP/substrate interface simultaneously, balancing the topological charge of the system and dissipating the accumulated strain. This suggests that manipulating the substrate geometry can control the formation of mesoscale topological defects in BCP networks. Our methodology centered on X-ray nanotomography opens new avenues to study the role of topological defects in soft matter and other 3D nanostructured systems.

Results

Three-dimensional imaging techniques are indispensable for directly observing and classifying defects in 3D nanostructured networks. Transmission electron microtomography (TEMT) has been successfully used to provide high-resolution 3D images[25-28]. However, the accessible sample volume is limited, particularly in thickness due to the electron mean free path (typically about 100 nm[29]). Alternating rounds of focused-ion beam (FIB) slicing and scanning electron microscope (SEM) imaging, known as FIB-SEM or slice-and-view SEM (SV-SEM)[30,31], affords an increased sample volume with a spatial resolution set by the depth of the FIB slices (as low as ca. 3 nm). However, FIB-SEM is a destructive technique, as the sample is progressively removed during imaging, meaning that subsequent measurements of the system's functionalities would not be possible. Here, we employed projection-based synchrotron X-ray nanotomography to non-destructively image and visualize a single diamond network with a monoclinic ($49\times75\times108$ nm$^3$) unit cell[24]. The network was formed by self-assembly via controlled slow drying (solvent vapour annealing, SVA[32-34]) of films of the triblock terpolymer ISG (polyisoprene-$b$-polystyrene-$b$-poly(glycidyl methacrylate), PI-$b$-PS-$b$-PGMA) initially swollen with tetrahydrofuran (THF) vapour. After forming the PI and PGMA phases into the inversion-paired networks of the alternating diamond morphology, the PI network was selectively removed by UV exposure and ethanol immersion, followed by backfilling the voided polymer template with gold by electrodeposition. Using ptychographic X-ray computed tomography (PXCT)[35,36], we produced a high-resolution 3D image of the entire volume of a cylindrical micropillar sample containing approximately 70,000 unit cells of the polymer-encased Au single-diamond network. In this study we benefit from recent instrumentation and algorithms advancements at the cSAXS beamline of the Swiss Light Source[37-40] that allowed us to achieve the high spatial resolution needed to resolve the unit cells of the lattice. Detailed descriptions of the sample preparation and PXCT experiment are provided in the Methods section and the Supplementary Information.

With 11.2 nm spatial resolution we were able to resolve individual unit cells and struts, analyze their connectedness and orientation in the network, quantify the distortions of the unit cells across the volume, and segment and render the domains. After data segmentation and rendering, we observe that the single diamond network comprises individual domains of different orientations separated by boundaries (Figure 1 and Extended Data Figure 1). More specifically, we identified: domain 1 (approximately 46,550 unit cells, color-coded in yellow), domain 2 (approximately 16,800 unit cells, color-coded in green), domain 3 (approximately 5,250 unit cells, color-coded in blue), micro domains (approximately 1,400 unit cells, color-coded in purple) that formed at the top of domain 2, and a domain boundary that extends throughout the fabricated sample. The cross-sectional patterns of the two largest domains (domain 1 and domain 2 in Figure 1b) are consistent with the single-diamond space group, in which the nodes are connected to four neighbors. We described the detailed analysis of the exact domain geometries in a separate publication[24]. Both domains have the <110> crystallographic direction out-of-plane and are rotated in-plane relative to each other by approximately 50°. Extracted subvolumes from these two domains (Figure 1c,d) can be mapped onto the single-diamond level-set model (see Figure 1e and ref.[24] for details). For the smallest domain in the fabricated sample (domain 3 in Figure 1b), it was impossible to identify the underlying structure of the domain unambiguously. Although this is most likely due to the formation of an imperfect bicontinous structure either during the drying of the terpolymer template or during the metal deposition, we cannot completely exclude imaging artefacts due to the position of the domain at the sample's edge. A detailed discussion of the domains' identification methodology can be found in the Supplementary Information.

The three domains are separated by a branched domain boundary, which is one to two unit cells wide (approximately spanning 100 nm, color-coded in red in Figure 1b and 2a). We identified two points of interest along this domain boundary as it travels from the edge of the fabricated sample: (i) a kink in the direction of the boundary between domain 1 and domain 2 (marked with + sign in Figure 2b), and (ii) a branching where domain 1, domain 2, and domain 3 meet (marked with - sign in Figure 2b). To better understand these features of the domain boundary and the emergence of the three domains, we present a thorough investigation of the domains' geometry in the vicinity of the boundary including a detailed characterization of the overall domain topology based on a segmentation and 2D slicing of the high-resolution 3D tomogram.

First, we analyse the diamond unit cell orientation pattern within the domain structure's top surface (Figure 2). Two striking textures, with their centers separated by ca. 2.7 µm distance, are observed at the previously identified points of interest within the domain boundary. Morphologically, the collective orientational rearrangements give rise to one comet-like texture (ca. 0.4 µm by 1 µm in size) with the "tail" aligned approximately perpendicular to the domain boundary, and one trefoil-like texture (ca. 1 µm by 1 µm in size) with a 120° symmetry, as schematically shown in Figures 2d,e, respectively. Note that, from fast Fourier transform (FFT) analysis, the peaks in reciprocal space corresponding to the two textures are located on a contour of constant radius (Figures 2f,g) whereas those of the "defect-free" areas of domains 1 and 2 form a rectangular lattice corresponding to the (110) surface of its underlying approximately cubic symmetry (Figures 2h, i). The central areas of the defect patterns therefore exhibit higher symmetry than the rest of the diamond network. The preservation of the high symmetry is characteristic of topological structures ranging from vortex cores in hard and soft condensed matter systems[41] to the hypothetical cosmic strings[42]. It thus points to a topological nature of the textures observed in our study.

To further confirm our hypothesis, we assign a winding number to each texture by assessing the orientation changes of the struts of the diamond unit cells surrounding these defects. Note that the network is continuous across the domain boundaries and between the two topological defects, as the domains remain connected despite their different in-plane orientations. However, due to the discrete nature of the reconstruction, we define the winding number as the radial sum (instead of an integral) over the angle $\theta$ in the vicinity of the topological defect as $l = \sum_{\theta=0º}^{360º} \overline{\omega}(\theta, R)$, where $\overline{\omega}$ is the orientation of the structures averaged within the area $\Delta R$. The textures show fractional winding numbers of opposite signs with $l$ =+0.57 for the comet-like texture and $l$ =-0.51 for the trefoil-like texture (Figure 2i,j). The difference between the calculated values (+0.57 and -0.51) and the values expected for this kind of defects (+0.5 and -0.5) are due to local variations of the orientation likely caused by the top and bottom interfaces' influence. It is important to note that through ca. 230 nm thickness where the centers of topological defects can be traced, they move by only ca. 160 nm laterally together with the domain wall. This and the relatively small variation of the orientation between the top and the bottom interfaces suggests a 2D character of the topological defects.

To understand the origins of the topological defects and how they are involved in the process of domain formation we analyze distortions in the single diamond network (see Figures 3 and 4). We start with a simple undistorted model structure based on single-diamond level-set equation. We observe complex patterns when the model structure is sliced at different angles relative to the vertical axis (see Extended Data Figure 2a). When the model structure is rotated, these patterns are altered (see Extended Data Figure 2b-d). It suggests that slicing the experimental data along vertical axis and analyzing distortion of the diamond cross-sectional patterns can be used to investigate distortions (that translate into strain) in the network. Slices

1-3 (Figure 3d) show cross-sections through the comet-like texture and then through the boundary between domains 1 (yellow domain in Figure 1) and 2 (green domain in Figure 1). At the texture itself, the single-diamond lattice shows distortions that propagate away from the domain boundary, appearing as a uniform bending of the diamond pattern in domain 1 and a rippling effect in domain 2. Although lattice strains are expected in solvent-annealed BCPs[32-34], these produce uniform compressive or shear distortions of the unit cell globally across the sample[24]. However, for the distortions observed here, the associated strain appears to decrease progressively with increasing distance from the defect along the boundary between domains 1 and 2 (as seen as progressively reduced bending of the diamond cross-sectioned patterns in Slices 1-3 in Figure 3d), suggesting that it is associated with the presence of a topological defect, in accordance with their conventional far-field behavior[43], rather than with a common domain boundary. Slices 5-7 show similar information for the trefoil-like defect, although the small size of domain 3 changes both the orientations and distances from the relevant defect of Slices 5-7 relative to Slices 1-3, making direct comparison difficult. Finally, Slices 4 and 8 show domains 1 and 2, respectively, at a distance from the textures and domain boundary equivalent to several tens of unit cells. The distortions observed in Slices 1-3 are absent in both cases.

Finally, we map distortion fields in the single-diamond domains by analysing the orientation of a selected cross-sectional pattern. The fringes in Figure 3d slice 5 are a compelling choice as their orientation can be unambiguously segmented and mapped. Figure 4b and c shows how a rotation of the structure based on distorted level-set model by 2º results in a significant change in the fringe orientation from parallel to the substrate (chosen as 0º reference) to an angle of 62º from the substrate. The significant change of the fringe orientation suggest high sensitivity to the distortions. We apply this approach to the experimental data for each domain (see Figures 4d,e) to qualitatively assess the deviations from the dominant (undistorted) domain structure. We observe that the largest domain (domain 1, color-coded as yellow in Figure 3a) has two areas with slightly different orientation of the fringes (near 0º and -26º rad) that translates into a sub-1º distortion. This, in turn, suggests a non-negligible difference in the unit cell orientations, indicating the presence of two regions with coherent distortions within a single domain. The most distorted areas (+60º and -60º fringe misorientation) with 2º unit-cell distortion meet at the comet-like topological texture. Moreover, Figure 4e shows that the distortion field has a depth profile with one of the distorted regions rolling onto the other. This observation of spatial localization of two distorted regions within a single domain allows us to infer the influence of the free surface of the single-diamond network and its interface with the substrate on the nucleation of the domains and topological textures.

From our observations we postulate that both topological defects emerge simultaneously, balancing the strain produced by impinging domains. The defects thus act as pinning points fixing the direction and positions of the boundaries between the domains. The comet-like texture emerges due to two mismatched domains while the trefoil texture forms at the intersection of three domains. This mechanism may emerge as a method of fine tuning the structural quality of the domains, as the comet-like topology preserves long-range order in its host domains, as seen in particular from domain 1. This structural synchronization and stress-reducing property of topological defects has been previously observed in biological systems[6-9] but are yet to be exploited in self-assembled artificial materials.

## Discussion

Our X-ray tomography study of a large volume (about 70,000 unit cells) of a self-assembled BCP single-diamond network backfilled with Au reveals pairs of extended topological entities with subtle orientational changes over several tens of unit cells. The observed defects are cross-domain entities, centered on the domain boundary yet belonging to several domains. This differs substantially from the topological defects within individual domains previously observed in other self-assembled 3D networks[30,31,44]. At first glance, the comet-like and trefoil-like textures studied here morphologically resemble topological defects that occur in diblock co-polymers such as PS-*b*-PMMA, which form into cylinders parallel to the substrate or substrate-perpendicular lamellae[45,46]. However, in such systems the defects are abrupt disclinations that break across quasi-2D stripe patterns, rather than collective orientational changes.

Instead, the micron-scale comet and trefoil in the BCP-templated single-diamond network have more in common with the analogous topological textures known from studies of nematic liquid crystals[2,46] and recently also observed in *Hydra* single-celled organisms[6] and colonies of motile bacteria[7,8]. In contrast, the disclinations in cylinder-forming and lamella-forming BCPs in biological systems more closely resemble the (millimeter-scale) +1/2 and -1/2 defects in the skin on the fingertips, occuring in topological charge-conserving pairs[47]. In all the previously reported non-BCP systems, order is governed by a director, a modulo-$\pi$ vector defined by the long axes of the rod-shaped bacteria in the colonies, the preferential local order axis of nematic molecules in liquid crystals, actin filaments in the single-celled organism *Hydra* and the ridges in fingerprints. Our work identifying analogous defects in BCP diamonds suggests that a director may also exist in this system, in the form of the strain field revealed by comparing experimental patterns with undistorted patterns from a level-set diamond model. Remarkably, topological defects of the observed type can exist in the diamond network, since similar self-assembled 3D networks in biological samples rather exhibit dislocations[44], analogous to "hard matter" atomic crystals.

The methodology devised in our study opens the pathway to future discoveries of topological defects in self-assembled 3D networks, both natural and synthetic. Moreover, the 3D fringe orientation analysis used in our work can be applied to other 3D periodic structures to investigate the role of displacement fields in the system where analytical modelling is challenging. The achieved high 3D spatial resolution combined with large sample volumes can be used to study defect formation over macroscopic distances in self-assembly processes. Understanding defects naturally leads to the possibility of defect engineering to realize new phenomena and improve materials' properties. Combined with its unique capability to control self-assembly of complex 3D nanostructures at a scale currently inaccessible by conventional lithographic fabrication, soft matter with controllable defects opens an attractive new route for next-generation nanotechnology.

More fundamentally, extended topological defects in the self-assembled system observed in our work share characteristics of synthetic (*e.g.* liquid crystals) and biological systems. This supports the emerging consensus that topology-driven physics is one of the leading mechanisms for structure formation in soft matter[48], and further suggests that self-assembly can be used as a model process to investigate the role of topology in nature.

Materials and methods

*Sample preparation*

Films of a nanostructured 3D diamond morphology were prepared as detailed in the Supplementary Information by self-assembly of a polyisoprene-*b*-polystyrene-*b*- poly(glycidyl methacrylate) (PI-*b*-PS-*b*-PGMA, ISG) triblock terpolymer under solvent vapor annealing (SVA). The total ISG number average molar mass, $M_n$, was 67.4 kg/mol with a polydispersity index of 1.08, and volume fractions were $f_{PI}$= 0.29, $f_{PS}$ = 0.52, and $f_{PGMA}$ = 0.19, as described elsewhere[24]. The resulting PI diamond minority network was selectively removed from the ISG terpolymer film by UV exposure and ethanol immersion, followed by backfilling of the voided network space with gold by electrodeposition. The polymer-encased Au single-diamond film thus formed was around 600 nm thick, and its unit cell exhibits a monoclinic lattice distortion[24]. To study the nanoscale morphology of the diamond network, we extracted a pillar of 8 μm diameter containing the single-diamond layer from the substrate and transferred it onto a custom sample mounting pin (see Supplementary Information and Extended Data Figure 3).

*Data acquisition and tomographic reconstruction*
The experiment was performed at the cSAXS beamline of the Swiss Light Source using the flexible tOMography Nano Imaging end-station (flOMNI, see Extended Data Figure 4) [37,38]. The 2D projections of the sample were obtained using X-ray ptychography[36]. In this imaging technique, the sample is raster scanned with a monochromatic and spatially coherent focused X-ray beam in a manner which guarantees partial overlap between adjacent illuminated areas. The beam scatters from each scanned point on the sample and the resulting diffraction patterns are collected in transmission in the far field by an area detector. The recorded diffraction patterns are then iteratively inverted into real space images with quantitative phase and absorption contrast using phase retrieval algorithms[36,49]. Instrumental characteristics such as the detector noise and vibrations of the sample will contribute to loss of resolution, as in other imaging techniques. However, in X-ray ptychography the absence of the physical imaging lens allows recovery of high-resolution information that is aberration free and in theory is only limited by the X-ray wavelength and the acceptance angle of the detector, rather than the size of the illumination or scanning steps.

2D ptychographic images acquired at different rotation angles of the sample are used for the final tomographic reconstruction of the sample's volume. This extension of X-ray ptychography to 3D is termed ptychographic X-ray computed tomography (PXCT)[35,36]. In recent years PXCT found diverse applications such as non-destructive imaging of integrated circuits[37,38], biophotonic structures[39], and magnetization structures[40], achieving a resolution in 3D of 14.6 nm[38]. In our measurements, we demonstrated that PXCT can achieve uniform 3D spatial resolution of 11.2 nm (see Extended Data Figure 5), sufficient to visualize the structure of the individual single-diamond unit cells. To reconstruct the volume of the 8 μm diameter, 600 nm thick Au single-diamond layer in the sample, we acquired 2400 projections of the sample equally spaced over an angular range of 180º. The measurement of a single projection took about 50 s and resulted in 35 hours for the high-resolution tomogram. More details about the PXCT measurement and the real-space reconstruction of the Au single-diamond network in 3D may be found in the Supplementary Information.

Data availability
The data is permanently deposited at cSAXS repository and is available upon request from corresponding author.

Acknowledgements
This study was financially supported by the Swiss National Science Foundation (SNSF) (163220, 188647, 168223, 190467), the National Center of Competence in Research *Bio-Inspired Materials* (51NF40-182881), and the Adolphe Merkle Foundation. This project had also received funding from the European Union's Horizon 2020 research and innovation programme under the Marie Sklodowska-Curie grant agreement no. 706329/cOMPoSe (I.G.). This work was also funded under grant agreement no. 731019/EUSMI (J.L., D.K.) and made use of the Cornell Center for Materials Research Shared Facilities supported by the NSF MRSEC program (DMR-1719875). U.W. thanks the National Science Foundation (DMR-1707836) for financial support. D.K. acknowledges funding from SNSF under grant no. 200021_175905. J.L. and S.F. acknowledge support from the Japan Society for the Pro motion of Science (JSPS) under KAKENHI 21K04816 and 19H05622, Cooperative Research Projects of CSIS, Tohoku University, and the Graduate Program for Spintronics (GP-Spin), Tohoku University. C.D. acknowledges support from the Max Planck Society Lise Meitner Excellence Program. The authors further acknowledge the Paul Scherrer Institut, Villigen, Switzerland for provision of synchrotron radiation beamtime at beamline X12SA (cSAXS) of the Swiss Light Source.

Competing interests
The authors declare no conflict of interest.

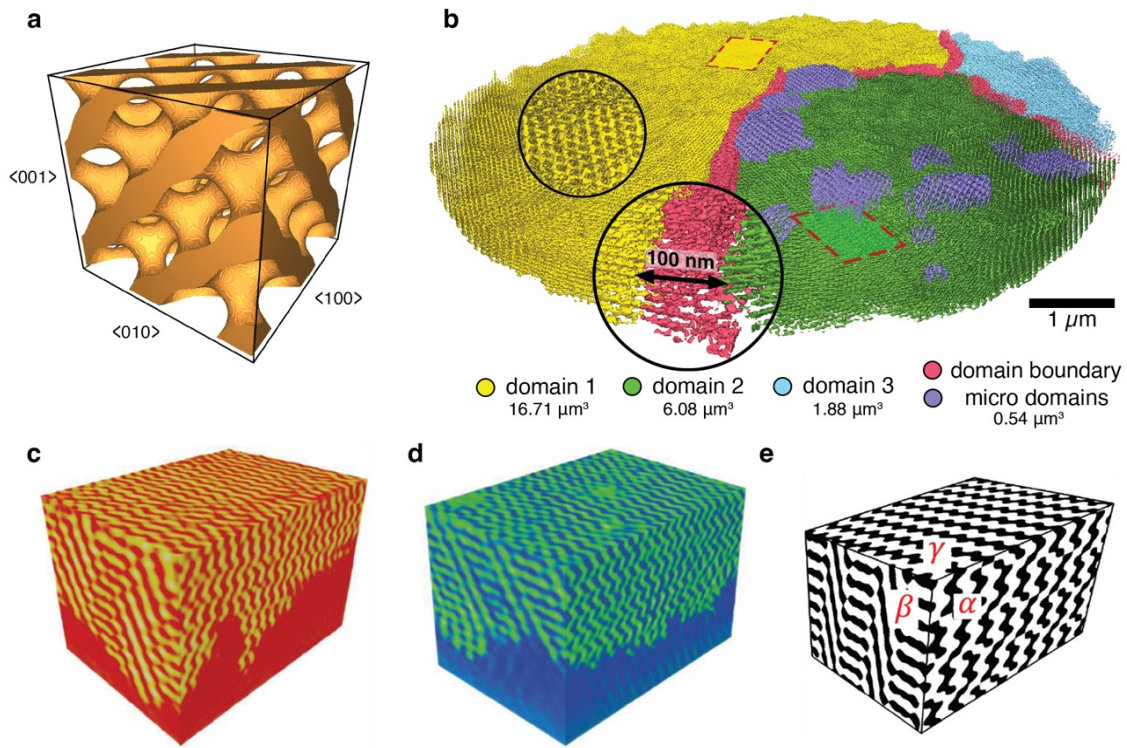

**Figure 1: Volume rendering and identification of (multi)domain structure.** (a) Three-dimensional rendering of the model structure (2x2x2 unit cells) determined from the level-set equations of the single diamond. (b) 3D rendering of the three domains spanning the largest part of the single-diamond network, isolated microdomains near the top surface, and the domain boundary. (c) and (d) Extracted renderings from domain 1 (yellow) and domain 2 (green), respectively, from the regions marked with dashed lines in (b). The two domains are oriented at 50° from each other. (e) Matching sub-volume found in the level-set generated single diamond (647 by 617 by 889 nm³). The angles between the planes are $\alpha = 92°$, $\beta = 77°$, $\gamma = 96°$. The top plane normal vector is (0.71, 0.70, 0.03) i.e., approximately <110>.

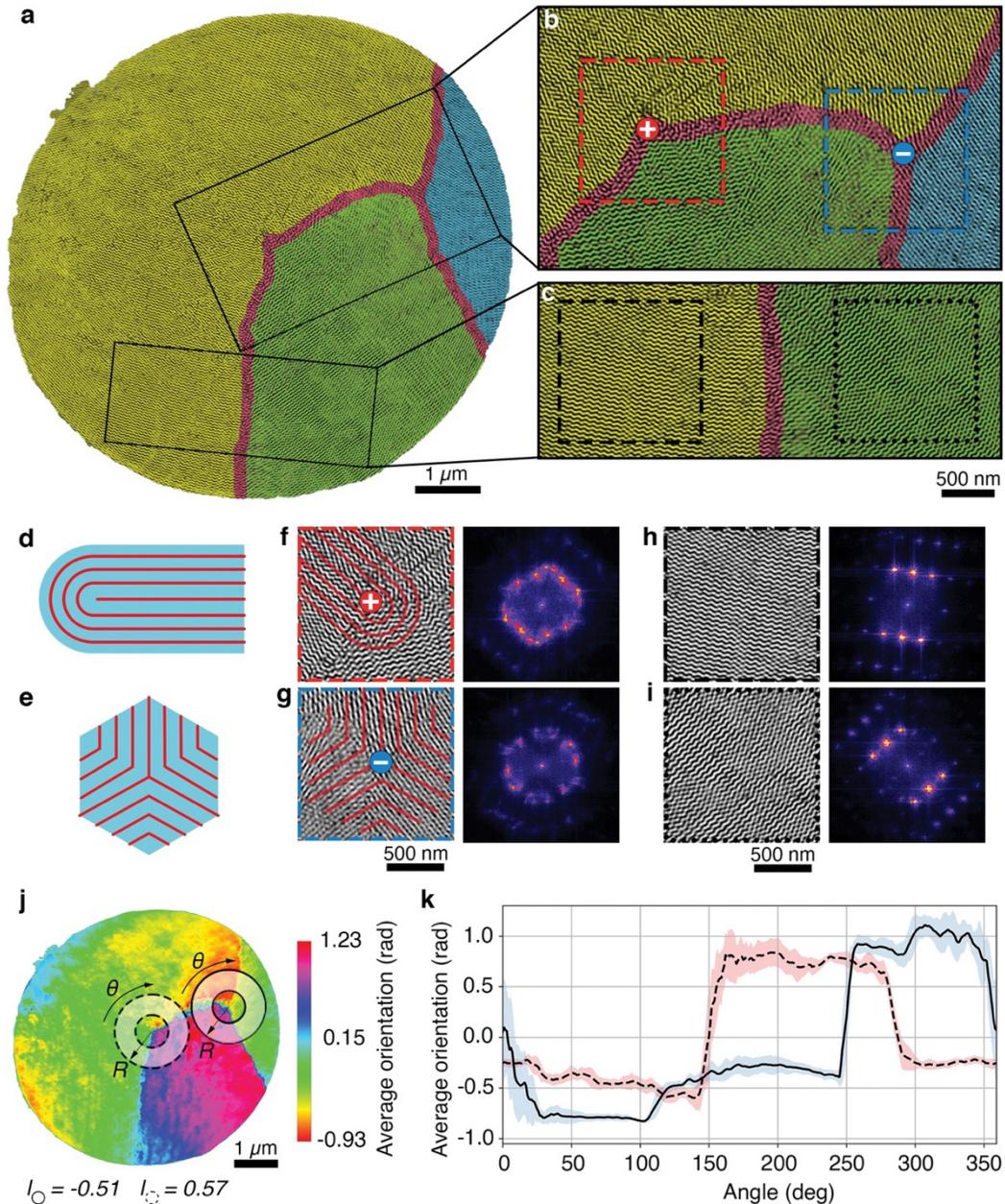

**Figure 2: Identification of topological defects.** (a) 2D slice through the sample's volume with colour-coded domains and the domain boundary as well as marked zoom-in regions in the panels ***b-c***. (b) Magnified region with two points of interest on the domain boundary marked with + and – signs. (c) Magnified region away from the points of interest. (d) Schematic illustration of a comet-like topological texture in ***f***. (e) Schematic illustration of a trefoil-like topological texture in ***g***. (f) View of highlighted area (red dashed line) in ***b*** with inferred topological texture and corresponding fast Fourier transform (FFT). (g) View of highlighted area (blue dashed line) in ***b*** with inferred topological texture and corresponding FFT. (h) View of highlighted area (dashed line) in ***c*** and corresponding FFT. (i) View of highlighted area (dotted line) in ***c*** and corresponding FFT. (j) Struts orientation map averaged over 26 slices of the tomogram, corresponding to 157 nm. Dashed-line circle marks comet-like topological defect with winding number 0.57. Continuous-line circle marks trefoil-like topological defect with winding number -0.51. (k) Angular dependence of the average orientation along the line plots marked by circled areas in ***j***. The shading on the graph indicates one standard deviation error.

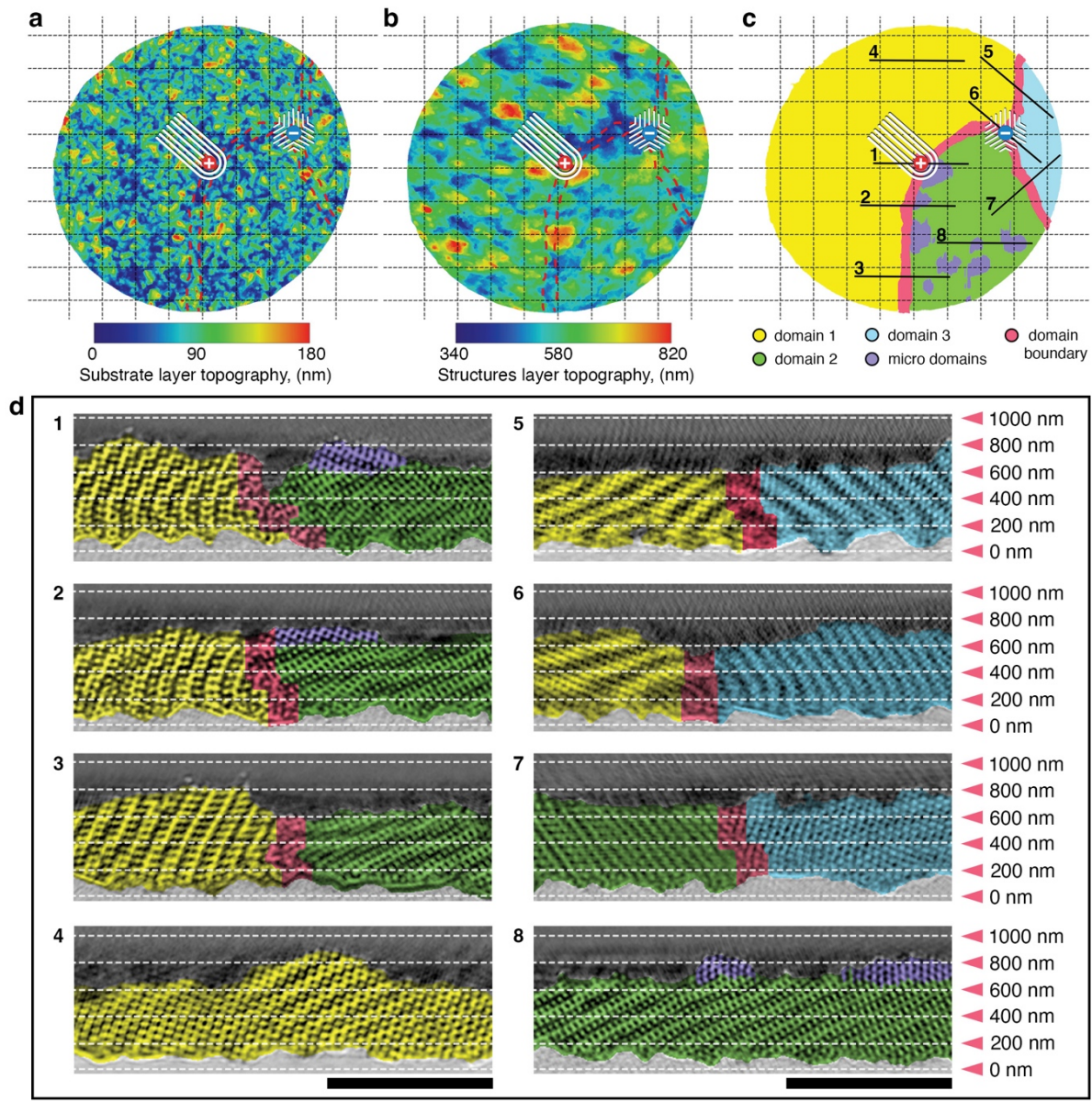

**Figure 3: Visualisation of the topological defects inside the 3D diamond network.** (a) Topography of the FTO-coated glass substrate with marks for the locations of topological texture and the domain boundaries. (b) Topography of the top surface of the diamond network layer with marks for the locations of topological textures and the domain boundaries. (c) Schematic of the domains, domain boundary, and the topological textures with lines indicating the location from where perpendicular views were extracted for panel *d*. (d) Cross-sections through the reconstructed 3D diamond network corresponding to the lines marked in panel *c*, showing Moiré fringes. The scale bar is 500 nm.

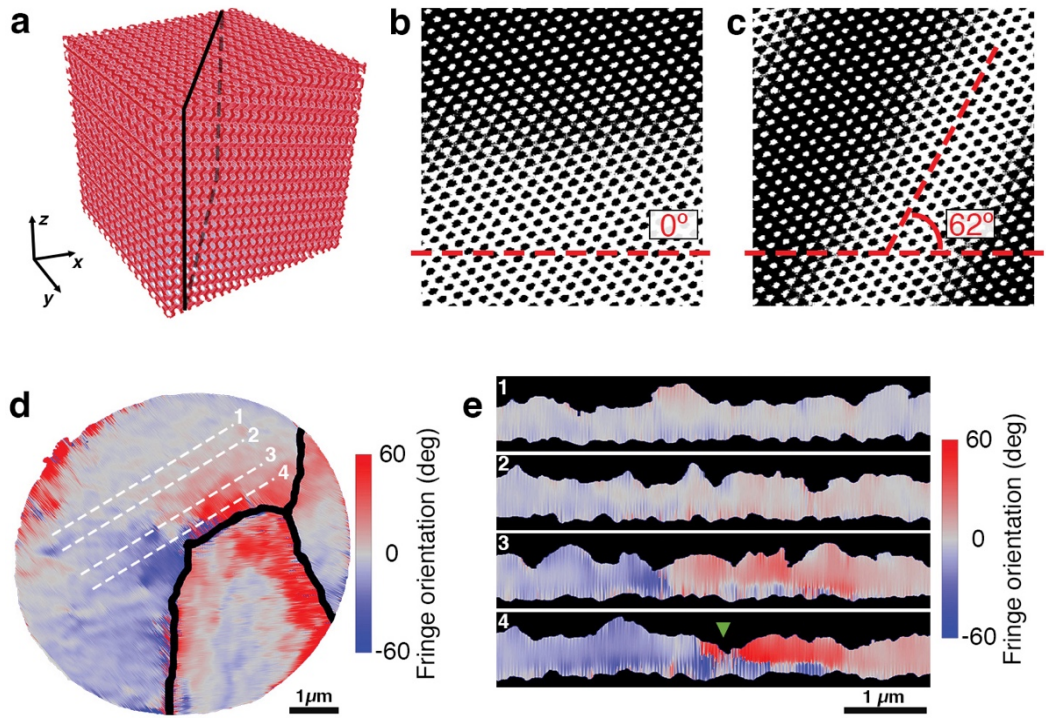

**Figure 4: Strain mapping.** (a) Three-dimensional rendering of a model structure. (b) Cross section through the model structure shown in *a* with a fringe at 0 degrees orientation. (c) Cross section through the model structure that has been rotated by 2 degrees around z-axis. This results in fringe rotation by 62 degrees. (d) Map of fringe orientations in experimental data that allows mapping the strain. Dashed lines indicate locations of cross-sectional views in the panel *e*. (e) Cross-sections through the orientation map with lines marked in *d*. The triangle in subpanel 4 marks location of the tail of the comet-like texture.

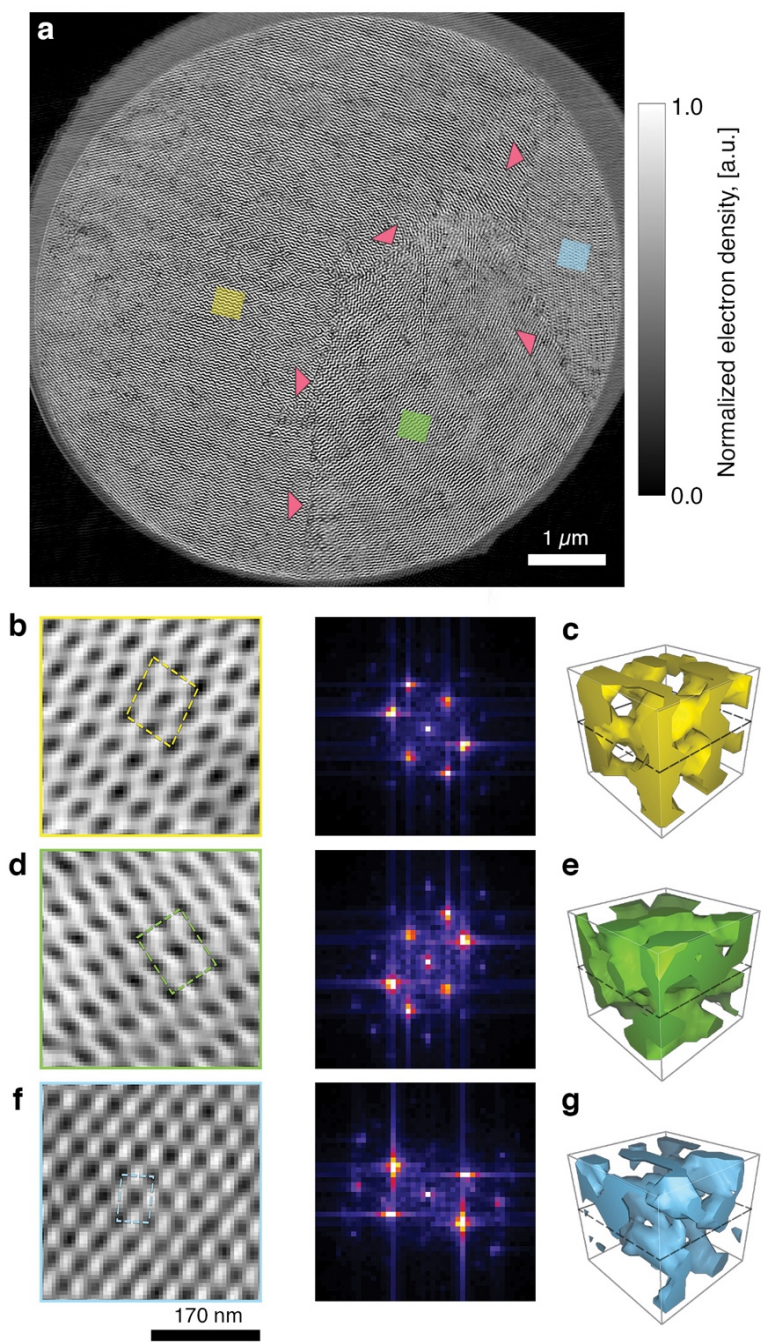

Extended Data Figure 1: (**a**) A slice through the volume (parallel to the substrate) indicating the domain boundary with triangular marks. (**b**) Slice through the yellow domain perpendicular to the sample surface and its Fourier transform. (**c**) 3D rendering of the small volume marked with a rectangle in **b**. (**d**) Slice through the green domain perpendicular to the sample surface and its Fourier transform. (**e**) 3D rendering of the small volume marked with a rectangle in **d**. (**f**) Slice through the blue domain perpendicular to the sample surface and its Fourier transform. (**g**) 3D rendering of the small volume marked with a rectangle in **f**.

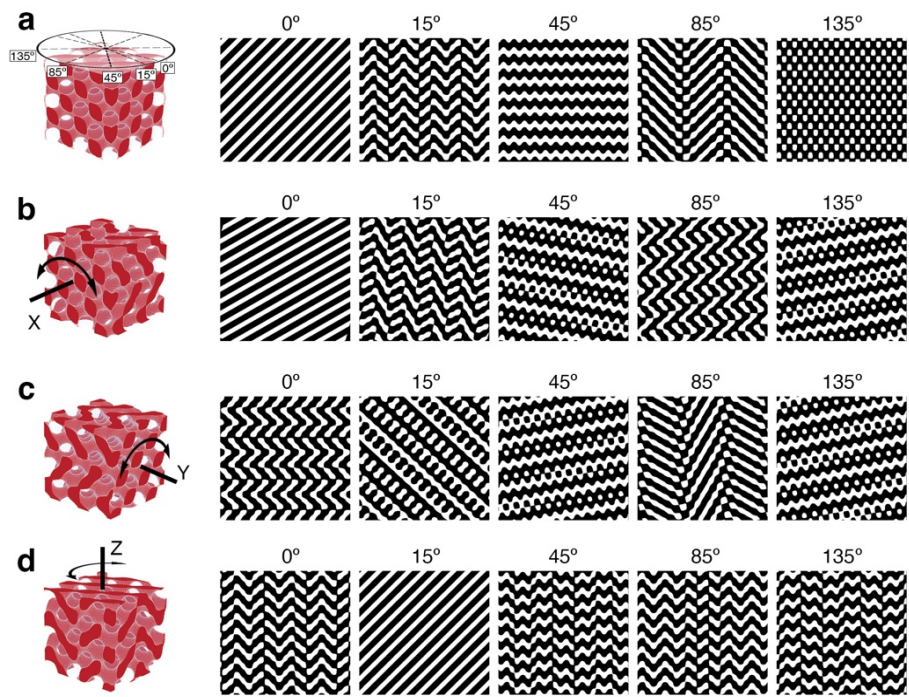

Extended Data Figure 2: (a) 3D rendering and slices through model diamond structure showing the different patterns. (b) 3D rendering and slices through the model rotated by 15º around the x-axis showing the dependence of the patterns on rotation. (c) 3D rendering and slices through the model rotated by 15º around the y-axis showing the dependence of the patterns on rotation. (d) 3D rendering and slices through the model rotated by 15º around the z-axis showing the dependence of the patterns on rotation.

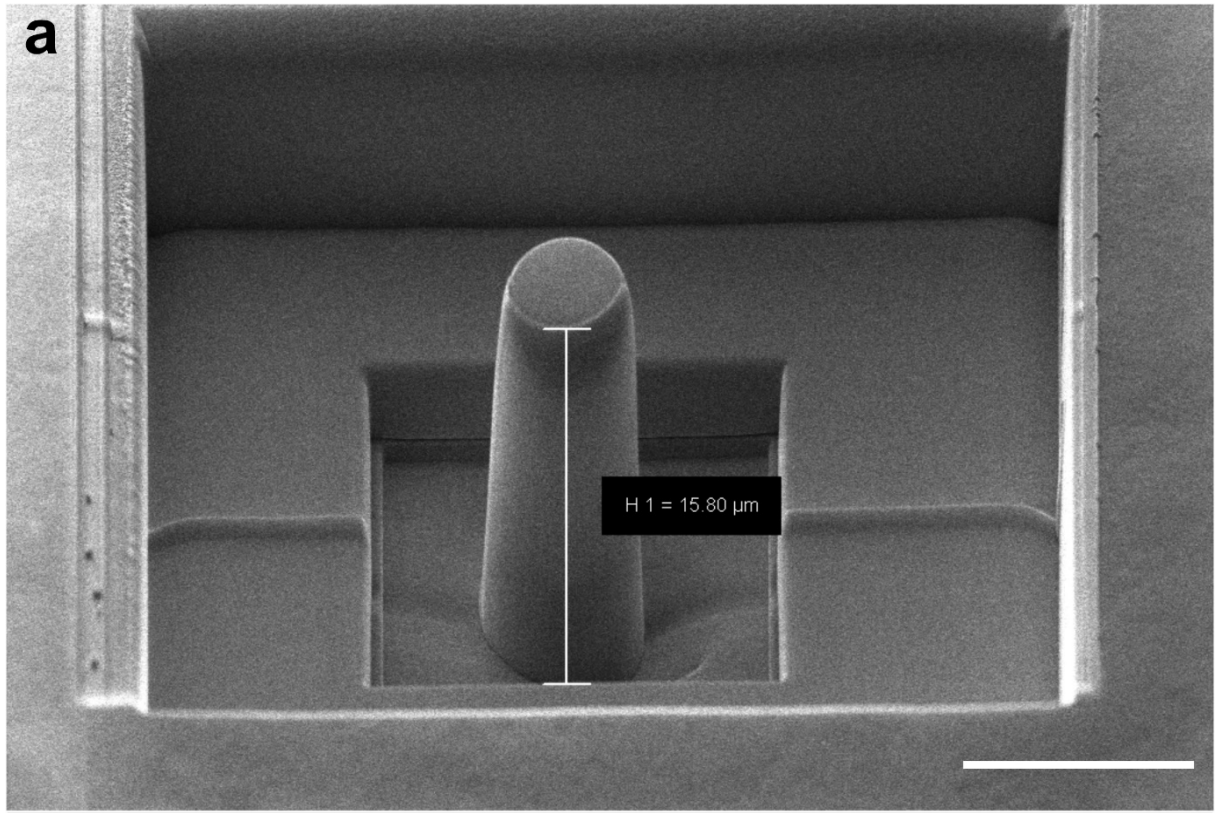
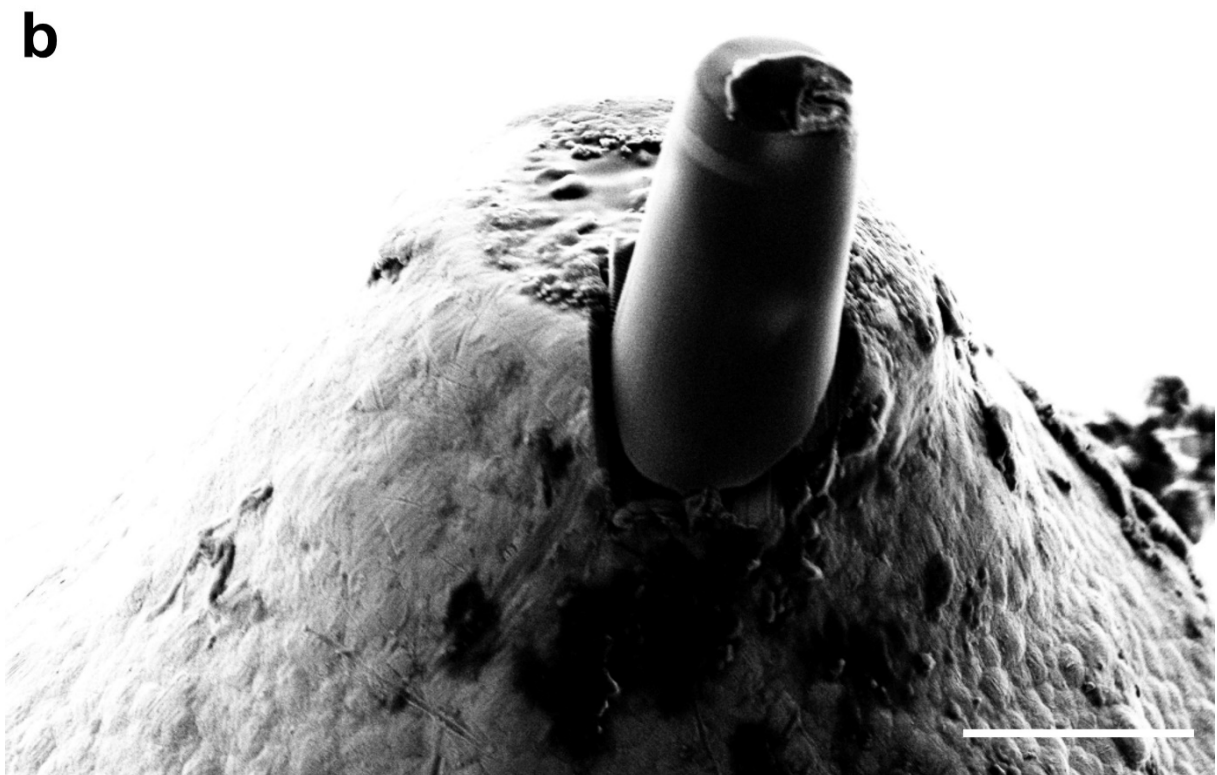
Extended Data Figure 3: Sample preparation by focused ion beam (FIB). (a) Scanning electron microscopy (SEM) image showing the sample after FIB milling of its surrounding material. (b) SEM image of the micropillar sample mounted on the OMNY pin. Scale bars are 10 µm.

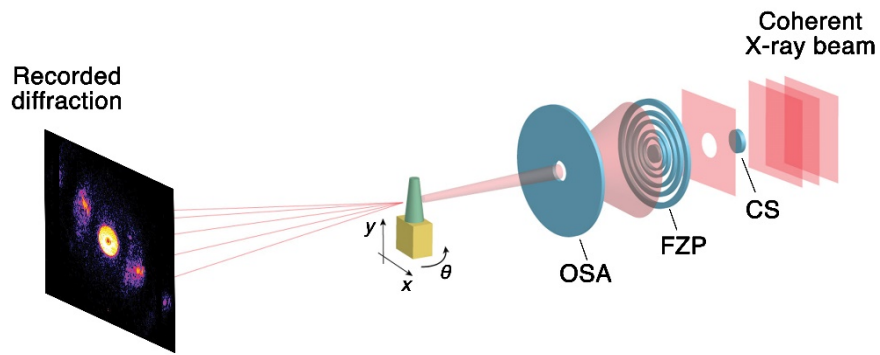

Extended Data Figure 4: Principal scheme of the ptychographic experiment. X-rays with a high degree of coherence are focused by a Fresnel zone plate (FZP) in combination with a central stop (CS) and an order sorting aperture (OSA) to define a coherent illumination onto the sample. X-rays propagate through the sample and are recorded by a 2D detector in the far field. The sample is scanned along x and y for ptychographic imaging, and then rotated in theta for tomographic acquisitions.

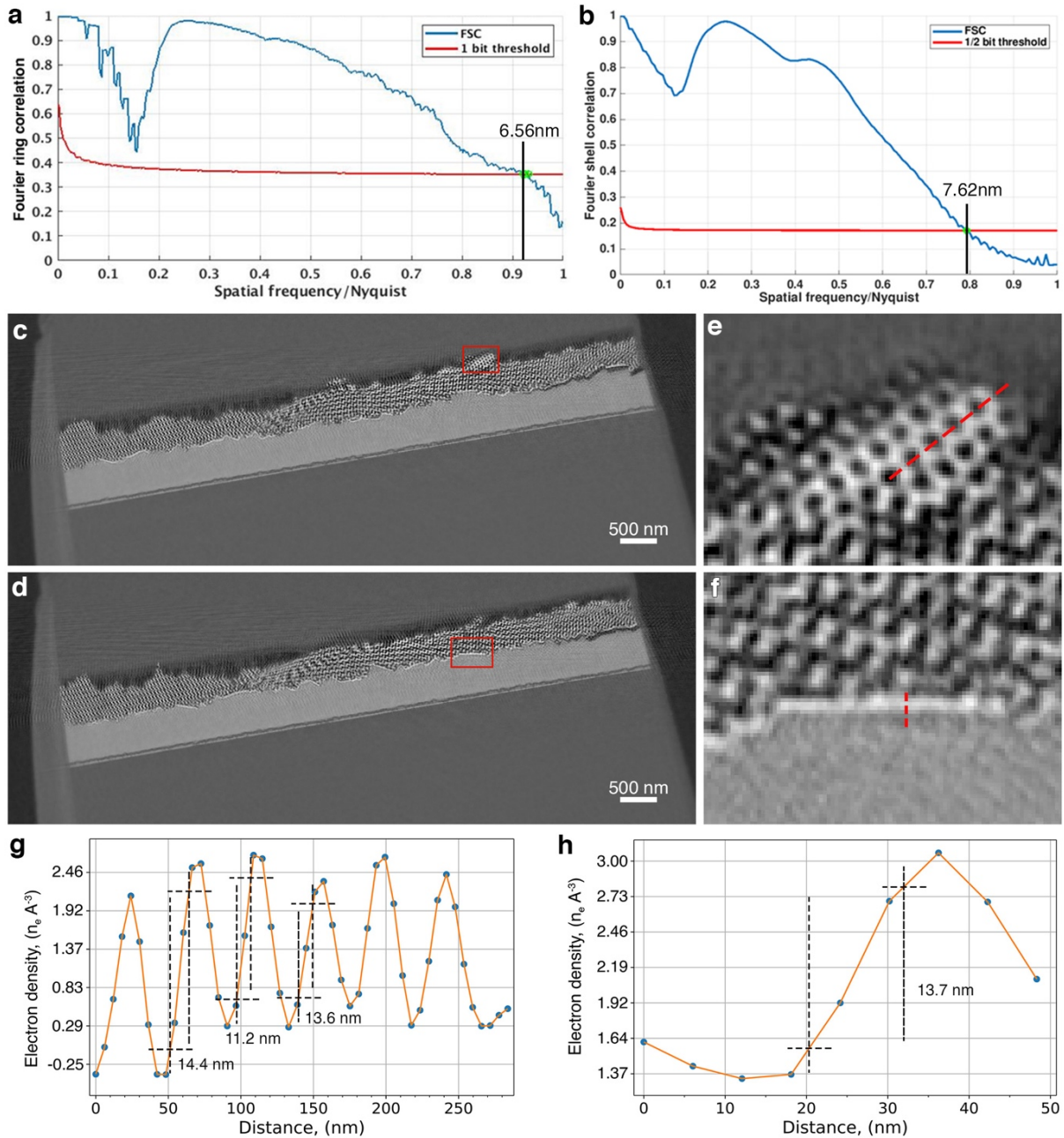

Extended Data Figure 5: Resolution estimation. (**a**) 2D resolution estimate of a single ptychographic projection. We show the Fourier ring correlation (FRC) between two ptychographic projections acquired at the same angle as a function of the spatial frequency. The point at which the FRC intersects the threshold curve calculated according to the 1-bit criterion is used as an estimation for the resolution in 2D corresponding to each of the images. (**b**) 3D resolution estimate of the tomogram. We show the Fourier shell correlation (FSC) between two sub-tomograms, each computed with half of the available projections, as a function of the spatial frequency. The point at which the FSC intersects a calculated threshold for the ½-bit criterion is an estimation of the 3D resolution of the full tomogram, using all the available projections. (**c-d**) 2D slices through the tomogram at different positions. (**e-f**) Magnified regions are indicated by red lines in (c-d), respectively. (**g-h**) Line profiles corresponding to the red dotted lines shown in (e-f), respectively. The numbers indicate the width of the profile line across sharp features of the image using the 10%-90% intensity criterion.

# Supplementary Information

## Contents

Block co-polymer self-assembly & Au replication of single-diamond network
Micropillar fabrication by focused ion beam (FIB)
Ptychographic and tomographic reconstruction
Resolution estimation
Structural stability of the sample to X-ray irradiation
Segmentation of domains
Substrate topography
Supplementary references
Supplementary figures

### Block co-polymer self-assembly & Au replication of single-diamond network

Single-diamond films were prepared from Au-filled polymer templates fabricated on fluorine-doped tin oxide (FTO)-coated glass substrates. Polyisoprene-*b*-polystyrene-*b*-poly(glycidyl methacrylate) (PI-*b*-PS-*b*-PGMA) triblock terpolymer (ISG) was synthesized by anionic polymerization as described elsewhere[1]. Before ISG film processing, FTO-coated glass was etched in piranha solution and subsequently silanized by immersion in a 4.3 mM solution (0.2% v/v) of octyltrichlorosilane (Sigma-Aldrich) in anhydrous cyclohexane (Sigma-Aldrich) for 12 s. ISG films on silicon and FTO-coated glass substrates were prepared by spin-coating a 10 % solution of the ISG terpolymer in anhydrous anisole (Sigma-Alrich) for 60 s at 1200 rpm with an acceleration of 500 rpm/s.

ISG-coated substrates underwent solvent vapor annealing (SVA) using tetrahydrofuran (THF) for 44 h in a custom-made annealing chamber described elsewhere[2]. The solvent reservoir temperature was set using a controlled water bath (ThermoFisherARCTIC SC150-A10B Refrigerated Circulator) to 23 °C, similar to room temperature to avoid condensation of solvent in the chamber inlet lines, and the sample was set to a lower temperature of 21.3 °C using a Peltier heating stage inside the annealing chamber. Thin film interferometry was performed on ISG films atop silicon substrates assuming a refractive index for ISG of n = 1.5, giving an estimated ISG film thickness of approximately 600 nm.

To create polymer templates for metal electrodeposition, the PI phase in the ISG terpolymer films was degraded by UV irradiation for 15 min (15 W, peak emission at 254 nm, sample-to-light source distance 11 cm (Analytik Jena, UV bench lamp GZ-97605-05)) and subsequently removed by immersion of the film in ethanol (Sigma Aldrich) for 30 min. The voided ISG polymer templates were then filled with Au by electrodeposition, using a Metrohm AutoLab PGSTAT302N potentiostat and Au plating solution (ECF 60, Metalor) with 0.5 % v/v brightener (a 66.7 mM $As_2O_3$ (Sigma Aldrich) solution in deionized water, with KOH added to adjust the pH to about 14). A three-electrode cell was employed with the FTO-coated glass substrate as the working electrode, a Pt electrode tip (Metrohm) as the counter electrode, and a Ag/AgCl with KCl reference electrode (Metrohm). The first step of the electrodeposition consisted of the nucleation of the Au nanostructure on the conductive FTO layer. Au was electrodeposited by applying cyclic voltammetry with a potential range of -0.4 V to -1.15 V at a scan rate of 0.05 V/s, followed by constant potential growth at a potential of −0.756 V. The remaining PS and PGMA phases were not removed after electrodeposition to protect the Au single-diamond during subsequent processing.

### Micropillar fabrication by focused ion beam (FIB)

Single-diamond micropillar samples were extracted from the Au-filled ISG film to perform tomographic measurements. Each micropillar was shaped by FIB (Focused Ion Beam) processing using a Carl Zeiss Microscopy NVision40 FIB-SEM with Ga ion source. First, a layer of carbon ~1.5 μm thick was deposited on the surface of the Au-filled polymer film in the FIB sample chamber to prevent damage to the single-diamond layer by the ion beam. Next, a 30 kV/13 nA $Ga^+$ beam excavated a square area of the protection layer and underlying layers to ~12 μm depth to begin defining a pillar-shape (see Extended Data Figure3). The final shape and diameter (~8 μm) of the pillar-shaped samples measured in the study were refined by a 30 kV/3 nA $Ga^+$ beam, used to excavate to the final ~15 μm depth to minimize the possibility of beam damage to the pillar.

After FIB milling, each pillar sample was picked up by a Kleindiek NanoControl NC40 micromanipulator system integrated with the NVision40 FIB/SEM. The micromanipulator was fixed to the top of the pillar's carbon protection layer by depositing ~0.3 μm thick carbon patch and the bottom of the pillar was cut from the glass substrate by a 30 kV/150 pA $Ga^+$ beam. Then the pillar was picked up and fixed close to the tip of a custom OMNY pin[3] by further carbon deposition around its base (see Extended Data Figure 3).

*Ptychographic and tomographic reconstruction*

In our work we used X-ray photons with energy of 6.2 keV selected by a fixed-exit double crystal Si(111) monochromator. The defined illumination on the sample (flux of $3.3 \times 10^8$ photons/sec) was produced by a central stop (CS), a Fresnel zone plate (FZP) with 170 μm diameter and 60 nm outermost zone width, and an order sorting aperture (OSA) (Extended Data Figure 4). The FZP was fabricated with locally displaced zones, designed for an optimal illumination for ptychography[4]. The sample was mounted about 0.5 mm downstream of the optics on a piezo stage that allows accurate positioning with respect to the beam-defining optics via external laser interferometry[5]. At this position, the illumination on the sample had a diameter of about 2 μm. We used an Eiger 1.5M detector with 75 μm pixels positioned 2.264 m from the sample to record the scattered X-rays. The detector is mounted directly inside the vacuum flight tube to reduce air scattering contributing to the experimental noise. Each projection was acquired using a Fermat spiral ptychographic scan[6] with 0.5 μm step size, covering an area of 12.5 μm × 3.7 μm with an exposure time of 0.1 s per point. During scanning, a combined motion of the FZP and the sample was used to minimize the scanning time[7].

In X-ray ptychography the illumination on the sample needs to be larger than the step size between neighboring positions on the sample. The resulting overlap between neighboring scan areas can be used to enforce consistency in the iterative phase retrieval, resulting in a fast and reliable convergence[8,9]. To reconstruct high-resolution projections it is crucial to evaluate the position reached after the movement between scanning points and ensure that it does not fluctuate beyond the desired resolution within a single exposure. The flOMNI instrument was designed to resolve the discussed stability issue through external metrology[5]. This instrument uses differential laser interferometers to measure relative positions of the sample and optics stages. These measurements are closed in a feedback loop, allowing to actively adjust sample position with respect to the optics during acquisitions. This instrument undergoes continuous development and has proven capability of reaching 14.6 nm[10]. The most recent modification of the setup that helped to advance the resolution achieved in the current paper is an additional feedback loop that increases the stability of the wobble angle of the sample stage.

Diffraction patterns from the sample show clear Bragg peaks, confirming, prior to the reconstruction, the highly ordered nature of the sample (see Figure S1). To reconstruct the diffraction patterns, we used the PtychoShelves package developed by the X-ray Coherent Scattering group at the Paul Scherrer Institute[11] and a detector area of 1000 x 1000 pixels, which resulted in 6.04 nm pixel size in the reconstructed 2D images. To reconstruct each ptychographic projection, we used 500 iterations of least square maximum likelihood with compact set approach (LSQ-MLc) method[12]. After zero- and first-order phase correction and registration of the projections[13,14], filter back projection using a Ram Lak filter with a frequency cut off of 1 was used to perform the tomographic reconstruction. The reconstructed probe indirectly confirms the correct reconstruction (see Figure S2), as it shows the expected divergent beam produced by the custom-aberrated FZP[4] after a propagation distance of 0.5 mm from the focus.

### Resolution estimation

The 2D resolution of each phase image was estimated as 7 nm by Fourier ring correlation (FRC) using the 1-bit threshold criterion[15] (see Extended Data Figure 5a). To reconstruct the volumetric distribution of the electron density we followed the approach described in earlier work[13]. The final 3D resolution estimated by Fourier shell correlation (FSC) using the 1/2-bit threshold is 7.62 nm (see Extended Data Figure 5b). However, it is important to do another independent test of the resolution to verify the estimate of the FSC. We used the 10%-90% line profile criterion (see Extended Data Figure 5c-h) that shows resolution estimates in the range of 11 nm and 14 nm, depending on the structure chosen for the estimation. To stay conservative in our resolution estimation, we favor the lower value provided by the line profile method instead of the estimate found by the FSC method. Thus, the 3D resolution achieved in the current experiment was 11.2 nm.

### Structural stability of the sample to X-ray irradiation

Another requirement for achieving high resolution in PXCT depends on the ability of the sample to withstand extreme doses of X-ray radiation without suffering structural changes. Instead of continuous acquisition, the 2400 equally spaced angular projections are acquired as 8 sub-tomograms (each containing 300 projections), in a non-sequential angular interval sequence, as has been developed for time-resolved tomography[16]. This approach allows monitoring of radiation damage induced by the focused X-ray beam. Although each sub-tomogram has worse resolution compared to the complete tomogram, due to the lower number of projections, we can still compare them to evaluate structural changes that take place during acquisition. Over the duration of several complete PXCT scans of the entire pillar sample, we observed no changes (see Figure S3) with time/X-ray exposure (estimated for this measurement as $1 \times 10^{10}$ Gray) to the microstructure of the Au single-diamond. Gold is an element that provides efficient momentum transfer to incoming X-rays, resulting in efficient scattering, and also has high thermal and electron conductivity, which combines with the sample's network structure to make the Au single-diamond highly resistant to the effects of X-ray radiation.

The strong scattering and excellent radiation resistance of gold single-diamond structures observed in this study may provide a solution to the need for new types of X-ray optical components and coatings required to keep pace with the extreme increases in photon flux in fourth-generation synchrotrons, and raises new prospects for single-diamond-based photonic and plasmonic devices for X-ray applications.

### Segmentation of domains

To identify the domains, we first performed a semi-automatic segmentation. The pipeline for the segmentation workflow is shown in Figure S4. We start by using a threshold to segment the air in the tomogram and watershed algorithm to segment the substrate and top surface layers. These 3 masks are then manually refined and their negative mask defines the layer with the diamond structures.

We apply two independent filters to segment the layer with the diamond structures further: (i) FFT with the 50x50 nm$^2$ window, and (ii) the 2D gradient filter. The FFT filter is based on analysis of the reciprocal space, which indicates the orientation of the unit cell.

However, this filter is very sensitive to local inhomogeneities. The gradient filter on the other hand is sensitive only to global orientation. By combining the results of both procedures, we get the initial masks of the domains. These masks are then manually refined.

We observe 3 large domains (domain 1 with 16.71 µm$^3$ volume, domain 2 with 6.08 µm$^3$ volume, domain 3 with 1.88 µm$^3$ volume), small domains (with 0.54 um$^3$ volume) precipitated on the top of domain 2, and the domain wall (see Figure 1). Small domains (color coded in purple in Figure 1b) which are only one to two unit cells thick and around five to six unit cells (around 300-400 nm) across share the same (110) crystallographic direction out-of-plane as domain 2 (on which they precipitate), but exhibit abrupt changes of in-plane orientation in the absence of any discernible boundary region.

*Substrate topography*

Looking at the topography of the fluorine-doped tin oxide (FTO) coated glass substrate and the top surface of the single-diamond film we do not observe any significant correlation between the domain morphology and the topography of the FTO-coated substrate (Figure 3a-c). Note that the substrate surface roughness (about 93 nm RMS) corresponds to 1-2 unit-cells of the single diamond in topographic height. The single diamond lattice in most cases appears to accommodate any local deformations within one- or two-unit cells along the substrate normal, as is evident in the cross-sectional slices through the sample in Figure 3d and commonly observed for self-assembled BCP structures on rough surfaces and in thin films. On the other hand, the top surface roughness (about 229 nm RMS) corresponds to around five diamond unit cells in height. While its topographic inhomogeneities are fewer in number, they are much larger in spatial extent (Figure 3d). Either interface can act as a nucleation site, and both could exert influence on the forming single-diamond simultaneously, as potentially observed for the yellow domain.

Supplementary figures

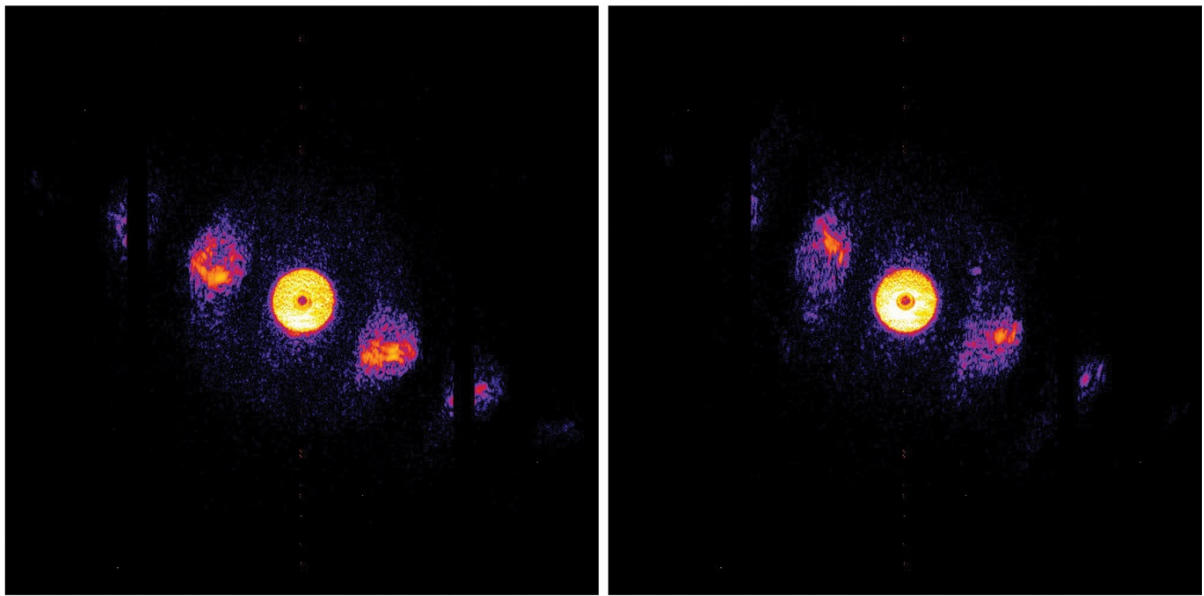

Figure S1: Examples of two diffraction patterns recorded at different positions on the sample.

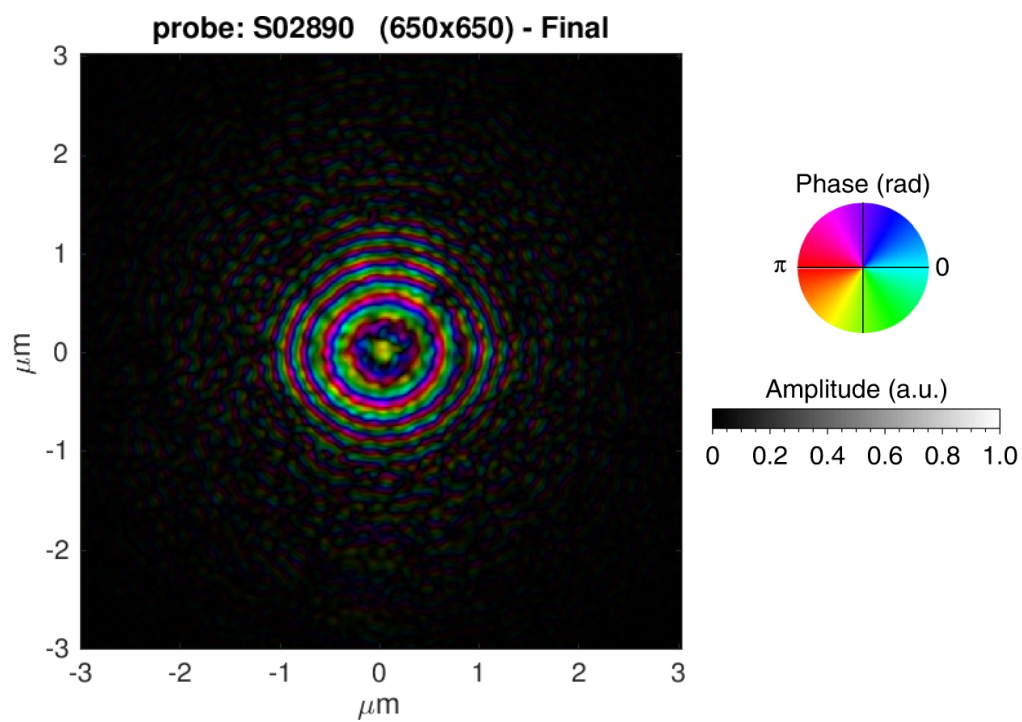

Figure S2: X-ray illumination on the sample reconstructed by ptychography at one angular projection. We show the complex-valued wave-front with amplitude and phase indicated by intensity and color, respectively.

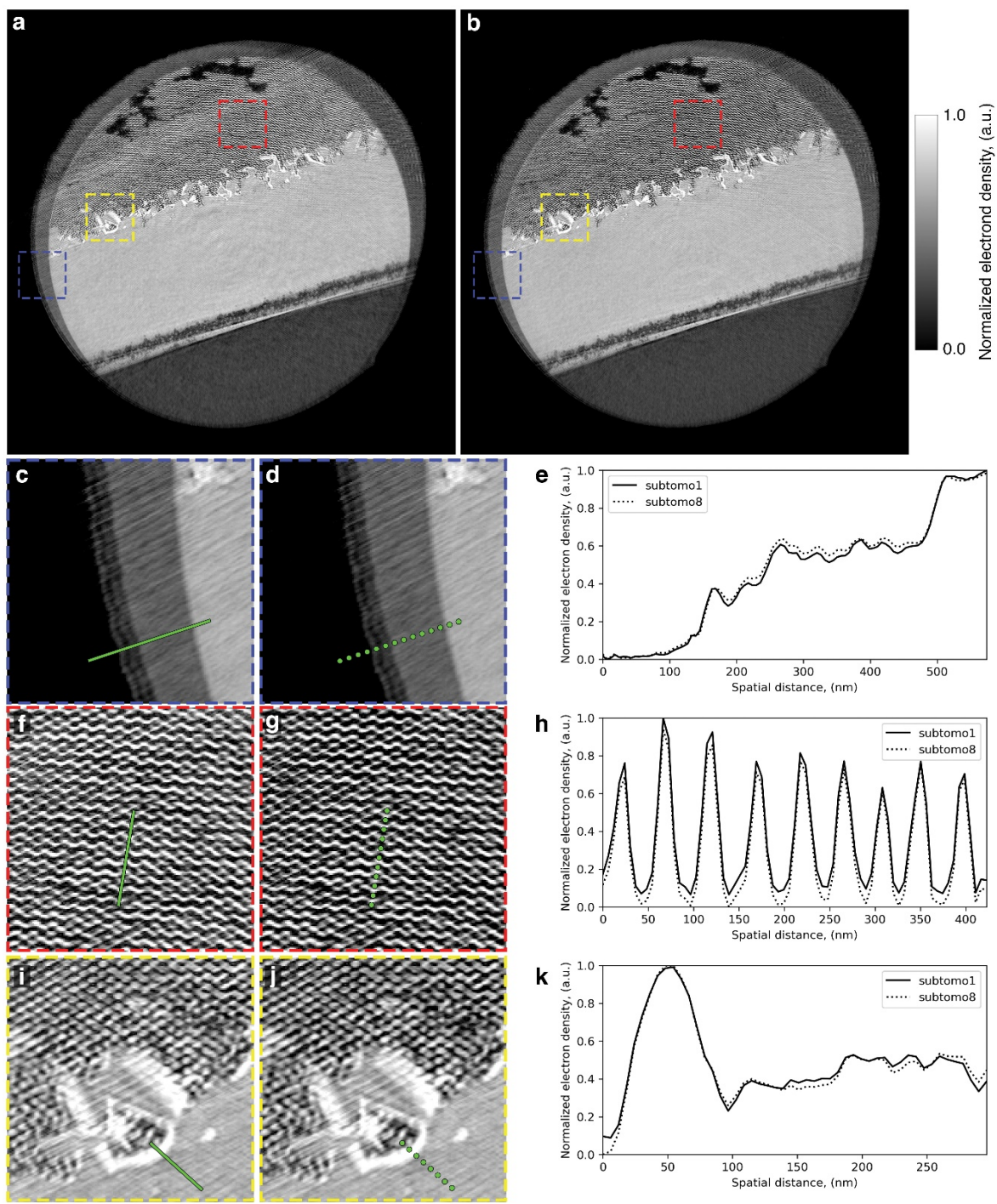

Figure S3: Evaluation of the sample change under prolonged exposure to the X-ray beam. (**a**) Slice through middle of the volume of the first sub-tomogram. (**b**) Slice through the middle of the volume of the last (8th) sub-tomogram. (**c**) Zoom into the region marked with blue square in **a**. (**d**) Zoom into the region marked with blue square in **b**. (**e**) Line profiles marked with lines in **c** and **d**. (**f**) Zoom into the region marked with red square in **a**. (**g**) Zoom into the region marked with red square in **b**. (**h**) Line profiles marked with lines in **f** and **g**. (**i**) Zoom into the region marked with yellow square in **a**. (**j**) Zoom into the region marked with yellow square in **b**. (**k**) Line profiles marked with lines in **i** and **j**.

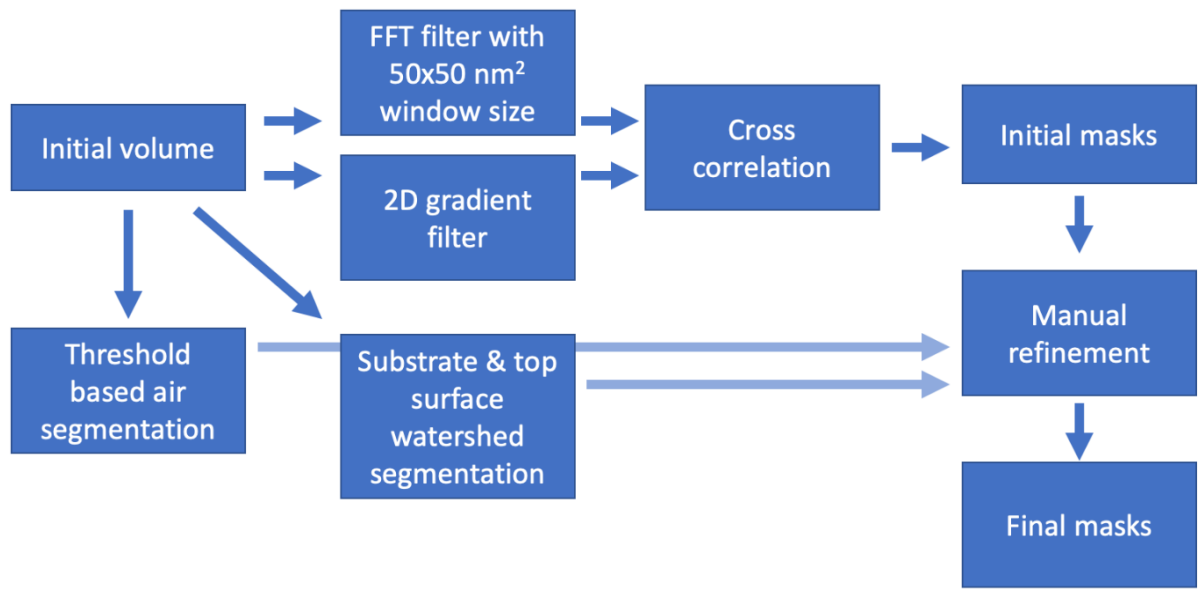

Figure S4: Workflow scheme for domain identification and segmentation. The initial volume refers to the 3D electron density distribution of the sample measured by X-ray nanotomography.